\documentclass[aps,prl,reprint,superscriptaddress,longbibliography,twocolumn]{revtex4-1}

\usepackage{amssymb}
\usepackage{amsmath}
\usepackage{graphicx}
\usepackage[colorlinks, linkcolor=blue, urlcolor=blue, anchorcolor=blue, citecolor=blue]{hyperref}
\usepackage{color}

\begin{document}
\title{Experimental realization of direct entangling gates between dual-type qubits}

\author{Chenxi Wang}
\thanks{These authors contribute equally to this work}%
\affiliation{Center for Quantum Information, Institute for Interdisciplinary Information Sciences, Tsinghua University, Beijing 100084, P. R. China}

\author{Chuanxin Huang}
\thanks{These authors contribute equally to this work}%
\affiliation{Center for Quantum Information, Institute for Interdisciplinary Information Sciences, Tsinghua University, Beijing 100084, P. R. China}

\author{Hongxuan Zhang}
\affiliation{Center for Quantum Information, Institute for Interdisciplinary Information Sciences, Tsinghua University, Beijing 100084, P. R. China}
\author{Hongyuan Hu}
\affiliation{Center for Quantum Information, Institute for Interdisciplinary Information Sciences, Tsinghua University, Beijing 100084, P. R. China}
\author{Zhichao Mao}
\affiliation{HYQ Co., Ltd., Beijing 100176, P. R. China}
\author{Panyu Hou}
\affiliation{Center for Quantum Information, Institute for Interdisciplinary Information Sciences, Tsinghua University, Beijing 100084, P. R. China}
\affiliation{Hefei National Laboratory, Hefei 230088, P. R. China}
\author{Yukai Wu}
\affiliation{Center for Quantum Information, Institute for Interdisciplinary Information Sciences, Tsinghua University, Beijing 100084, P. R. China}
\affiliation{Hefei National Laboratory, Hefei 230088, P. R. China}
\author{Zichao Zhou}
\email{zichaozhou@mail.tsinghua.edu.cn}
\affiliation{Center for Quantum Information, Institute for Interdisciplinary Information Sciences, Tsinghua University, Beijing 100084, P. R. China}
\affiliation{Hefei National Laboratory, Hefei 230088, P. R. China}
\author{Luming Duan}
\email{lmduan@tsinghua.edu.cn}
\affiliation{Center for Quantum Information, Institute for Interdisciplinary Information Sciences, Tsinghua University, Beijing 100084, P. R. China}
\affiliation{Hefei National Laboratory, Hefei 230088, P. R. China}
\affiliation{New Cornerstone Science Laboratory, Beijing 100084, PR China}
\date{\today}

\begin{abstract}
Dual-type qubits have become a promising way to suppress the crosstalk error of auxiliary operations in large-scale ion trap quantum computation. Here we demonstrate a direct entangling gate between dual-type qubits encoded in the $S_{1/2}$ and $D_{5/2}$ hyperfine manifolds of $^{137}\mathrm{Ba}^{+}$ ions. Our scheme is economic in the hardware, requiring only a single $532\,$nm laser system to entangle both qubit types by driving their Raman transitions. We achieve a Bell state fidelity of $96.3(4)\%$ for the dual-type Molmer-Sorensen gate between an $S$-$D$ ion pair, comparable to that for the same-type $S$-$S$ or $D$-$D$ gates. This technique can reduce the overhead for back-and-forth conversions between dual-type qubits in the quantum circuit with wide applications in quantum error correction and ion-photon quantum networks.
\end{abstract}

\maketitle

Trapped ions constitute one of the most promising physical platforms to build large-scale universal quantum computers \cite{doi:10.1063/1.5088164}. They support high-fidelity qubit state initialization, readout \cite{harty2014high,an2022high}, single-qubit gates \cite{harty2014high,brown2011single} and two-qubit gates \cite{PhysRevLett.117.060504,PhysRevLett.117.060505,clark2021high,loschnauer2024scalable}. Currently, all-to-all connectivity has been realized for tens of qubits in a one-dimensional ion chain \cite{chen2023benchmarking,Egan2021,PRXQuantum.5.030326} or in a quantum charge-coupled device (QCCD) architecture \cite{decross2024computational}. Global quantum manipulation has also been achieved for hundreds of qubits in a two-dimensional ion crystal \cite{guo2024site,britton2012engineered,bohnet2016quantum}.

Ions are subjected to heating from the environmental electric field noise and the collision with the background gas molecules \cite{wineland1998experimental,doi:10.1063/1.5088164}. Besides, ion shuttling in the QCCD architecture can result in additional heating \cite{wineland1998experimental,Kielpinski2002,decross2024computational}. As the motional modes of the ions heat up, the amplitudes and phases of the addressing laser beams feel stronger fluctuation, lowering the fidelity of quantum manipulation of the ionic qubits \cite{RevModPhys.75.281}. To solve this problem, sympathetic laser cooling is often required to dissipate the energy from the ion crystal in the run time through the random scattering of photons. Because such photons can destroy the quantum states of nearby ions, it is necessary that two different types of ions be used: one for encoding the quantum information and the other for the laser cooling cycle, such that the photons scattered by one type will not affect the other. In previous experiments, it is common to use two ion species \cite{PhysRevA.68.042302,guggemos2015sympathetic,decross2024computational} or two isotope ions \cite{PhysRevA.65.040304,PhysRevA.79.050305}. However, due to the different masses of the ions, they feel different trap frequencies in the same ion crystal, leading to large mismatch in their participation in the transverse collective phonon modes and hence low sympathetic cooling efficiency \cite{PhysRevA.103.012610}.

Recently, a dual-type qubit scheme has been proposed and demonstrated where, rather than using two ion species, the two types are encoded into two sets of energy levels of the same ion species like the $S_{1/2}$ and $F_{7/2}$ hyperfine clock states of ${}^{171}\mathrm{Yb}^+$ \cite{yang2022realizing,10.1063/5.0069544}. The two qubit types can be converted into each other coherently, allowing the switching of the qubit types on demand while maintaining high sympathetic cooling efficiency and low crosstalk error \cite{yang2022realizing}. Apart from laser cooling, this scheme can also be applied in suppressing the crosstalk from ancilla qubits during e.g. mid-circuit quantum state reset or detection \cite{yang2022realizing}, and the generation of ion-photon entanglement in a quantum network node \cite{Feng2024,lai2024realization}, similar to their dual-species counterparts \cite{doi:10.1126/science.1114375,PhysRevLett.99.120502,Negnevitsky2018,PhysRevLett.118.250502}. Despite these proof-of-principle demonstrations for the suppressed crosstalk between the dual types in various tasks, a crucial gadget for the future large-scale ion trap quantum computer remains to be realized, namely to entangle the dual-type qubits. This can enable wide applications like the non-demolition measurement of error syndromes in quantum error correction \cite{nielsen2000quantum} and the quantum teleportation between different computation modules in a distributed quantum computer \cite{RevModPhys.82.1209}.

\begin{figure*}[!tbp]
\includegraphics[width=\linewidth]{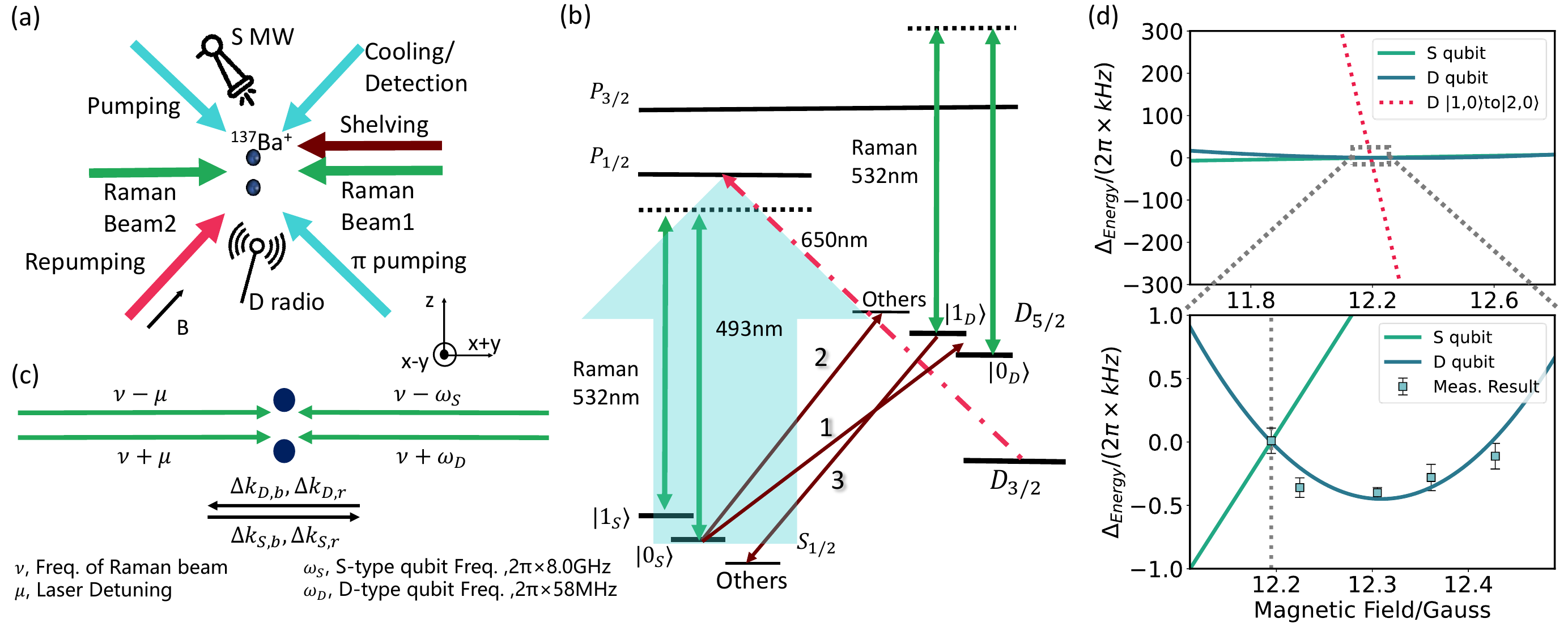}
\caption{Experimental scheme. (a) We trap two $^{137}\mathrm{Ba}^+$ ions in a blade trap along the axial $z$ direction.
We use global $493\,$nm laser beams (blue) for optical pumping, state detection, and Doppler and EIT cooling. A $650\,$nm laser (red) repumps the population in $D_{3/2}$ levels back to $S_{1/2}$ via the $P_{1/2}$ level.
A narrow-band global $1762\,$nm laser beam (brown) can connect the dual-type qubits encoded in the $S_{1/2}$ and $D_{5/2}$ manifolds. By tuning the laser frequency, this beam can also be used for the electron shelving of the $S$-qubit or the $D$-qubit.
A pair of focused $532\,$nm Raman laser beams (green) can drive Raman transitions on either type of the qubits. Besides, the $S$-qubit and the $D$-qubit can also be controlled by a microwave horn and a radio antenna, respectively.
(b) Relevant energy levels with laser beams represented by the same colors as (a). The $S$-qubit is encoded in $|0_{S}\rangle\equiv|S_{1/2},F=1,m_{F}=0\rangle$ and $|1_{S}\rangle\equiv|S_{1/2},F=2,m_{F}=0\rangle$, while the $D$-qubit is encoded in $|0_{D}\rangle\equiv|D_{5/2},F=2,m_{F}=1\rangle$ and $|1_{D}\rangle\equiv|D_{5/2},F=3,m_{F}=1\rangle$. The labels 1, 2 and 3 for the $1762\,$nm laser represents $D$-qubit initialization, $S$-qubit shelving and $D$-qubit shelving, respectively.
(c) Wave vectors and frequency components of the $532\,$nm laser beams for the Molmer-Sorensen gate. Each beam has two frequency components, driving Raman transitions for the $S$ and $D$ qubits simultaneously in a phase-sensitive geometry.
(d) The frequency shifts of the dual-type qubits versus the magnetic field. The $S$-qubit and the $D$-qubit have different hyperfine splittings. Here we subtract the qubit frequencies at the chosen magnetic field $B=12.2\,$G as the zero points. The upper panel compares the sensitivity of the $S$-qubit and the $D$-qubit with that of other typical Zeeman levels which do not form clock transitions (red dotted line). The lower panel shows a zoom-in near the sweet spot of the $D$-qubit, together with the experimental results.}
\label{fig:1}
\end{figure*}

In principle, quantum entanglement between dual-type qubits can be realized by that between same-type qubits sandwiched by the backward and forward qubit type conversions. This scheme is economic in the hardware, requiring only one laser system for entangling both qubit types, together with another one for qubit type conversion. In comparison, direct entangling gates between two qubit types generally require separated laser systems for individual types, similar to previous mixed-species entangling gates \cite{doi:10.1126/science.1114375,PhysRevLett.99.120502,Tan2015,Negnevitsky2018,Wan2019,Bruzewicz2019,PhysRevLett.125.080504}. However, the cost of the indirect scheme is the additional conversion steps whose fidelity is currently limited by many technical factors like the laser linewidth \cite{yang2022realizing,Feng2024}. It is thus desirable to combine the advantages of the two schemes to achieve the direct dual-type entangling gate using a hardware-economic laser system. Indeed, a commonly used $\sigma_z$-type entangling gate \cite{milburn2000,leibfried2003experimental,Lee_2005,PhysRevLett.117.060504} naturally supports the dual-type entangling gate if the hyperfine manifolds supporting the dual-type qubits (e.g. $S_{1/2}$ and $D_{5/2}$ of ${}^{88}\mathrm{Sr}$) can directly be coupled by suitable laser beams \cite{PhysRevA.107.022617}. Nevertheless, previous experiments only demonstrate entangling gates for same-type qubits, e.g. when both qubits are encoded in the $S_{1/2}$ manifold, the $D_{5/2}$, or the $S$-$D$ optical qubits \cite{PhysRevA.107.022617}. Here we demonstrate a more general scheme where a $\sigma_{\phi}$-type entangling gate between dual-type qubits can be realized using a single laser system via Raman transitions. We use $^{137}\mathrm{Ba}^{+}$ ions which have gained growing interest recently owing to their hardware-friendly transition frequencies in the visible range \cite{an2022high,huang2024electromagnetically,PhysRevA.81.052328,PhysRevResearch.2.033128}. We encode the dual-type qubits in the $S_{1/2}$ and $D_{5/2}$ hyperfine manifolds and demonstrate a direct dual-type entangling gate with a Bell state fidelity of $96.3\%$.


Our experimental setup is sketched in Fig.~\ref{fig:1}(a) with the relevant energy levels shown in Fig.~\ref{fig:1}(b). We confine two $^{137}\mathrm{Ba}^+$ ions in a blade trap with trap frequencies $(\omega_{x},\,\omega_{y},\,\omega_{z}) = 2\pi\times(1.6,\,1.7,\,0.2)\,$MHz. The ions can be cooled by Doppler cooling and electromagnetic-induced-transparency (EIT) cooling via $493\,$nm laser beams to near the ground state \cite{huang2024electromagnetically}.
We encode the dual-type qubits into $|0_{S}\rangle\equiv|S_{1/2},F=1,m_{F}=0\rangle$, $|1_{S}\rangle\equiv|S_{1/2},F=2,m_{F}=0\rangle$, $|0_{D}\rangle\equiv|D_{5/2},F=2,m_{F}=1\rangle$ and $|1_{D}\rangle\equiv|D_{5/2},F=3,m_{F}=1\rangle$ where the two types can be connected by the narrow-band $1762\,$nm laser with suitable frequencies. In the limit of a weak magnetic field, $m_F=0$ levels of $D_{5/2}$ form clock states with their transition frequencies insensitive to the perturbation in the magnetic field, analogous to the $|0_S\rangle$ and $|1_S\rangle$ states. However, under a finite magnetic field there will be a small linear dependence on the magnetic field, and since the hyperfine splitting for the $D_{5/2}$ levels is much smaller than that for the $S_{1/2}$ levels, this sensitivity to the magnetic field is also stronger. Therefore, as shown in Fig.~\ref{fig:1}(d), we set the magnetic field at $B=12.2\,$G, close to the sweet spot of $12.3\,$G where the transition frequency between $|0_D\rangle$ and $|1_D\rangle$ is immune to the first order change in the magnetic field. As we can see, at the chosen point, the two qubit frequencies are much more insensitive to the magnetic field (about $2\pi\times 12\,$kHz/G for the S-type and $2\pi\times 7\,$kHz/G for the $D$-type) than other Zeeman levels like $|D_{5/2},F=1,m_{F}=0\rangle$ and $|D_{5/2},F=2,m_{F}=0\rangle$ (red dotted curve) which can be probed by the radio antenna as shown in Fig.~\ref{fig:1}(a). More details about the choice of the magnetic field can be found in Supplemental Material.

The ions can be initialized into $|0_{S}\rangle$ with high fidelity by a $493\,$nm laser beam depleting the population in the $|S_{1/2},F=2\rangle$ states and microwave pulses depleting $|S_{1/2},F=1,m_F=\pm 1\rangle$ states, together with a $650\,$nm repumping laser for the $D_{3/2}$ states \cite{an2022high}. To initialize a $D$-type qubit in $|0_D\rangle$, we first prepare $|0_S\rangle$ and then convert it into $|0_D\rangle$ along the path labeled as 1 in Fig.~\ref{fig:1}(b). To increase the state preparation fidelity, we further verify that the target ion is dark under the $493\,$nm detection laser and the $650\,$nm repumping laser \cite{yang2022realizing}. Otherwise the data are discarded and the ions are reinitialized. For the measurement of the qubits, we use electron shelving to improve the fidelity \cite{RevModPhys.75.281}. Specifically, for the $S$-type qubit, we use $1762\,$nm $\pi$ pulses [labeled as 2 in Fig.~\ref{fig:1}(b)] to sequentially transfer the population in $|0_S\rangle$ into four hyperfine levels of $D_{5/2}$ (multiple pulses are used to suppress the imperfection of the $\pi$ pulse). For the $D$-type qubit, we combine $1762\,$nm $\pi$ pulses [labeled as 3 in Fig.~\ref{fig:1}(b)] with $493\,$nm pulses to transfer $|1_D\rangle$ into the $S_{1/2}$ and $D_{3/2}$ states which are bright under the detection lasers. We repeat the conversion for twenty times to suppress the imperfection in the $\pi$ pulses. Altogether, we obtain a state-preparation-and-measurement (SPAM) error below $0.3\%$ for both qubit types. More details can be found in Supplemental Material.

We use counter-propagating $532\,$nm laser beams to drive the Raman transitions for the $S$-type and $D$-type qubits, and further to realize the same-type ($S$-$S$ and $D$-$D$) or dual-type ($S$-$D$) entangling gates. As shown in Fig.~\ref{fig:1}(c), each of the counter-propagating laser beams has two frequency components. They form Raman transitions at the frequencies of $\omega_S\pm\mu$ and $\omega_D\pm\mu$, generating spin-dependent forces on the $S$-type and $D$-type qubits simultaneously in a phase-sensitive geometry \cite{Lee_2005}, thereby supporting the Molmer-Sorensen gates for both types of qubits. The frequency shift around $\omega_S=2\pi\times 8\,$GHz is created by an electro-optic modulator (EOM), and the frequency shift around $\omega_D=2\pi\times 58\,$MHz is introduced as the frequency difference of two acousto-optic modulators (AOMs) for the two Raman beams. Apart from the four desired frequency components, the EOM and AOM also generate other sidebands like $\omega_S+\omega_D$. Such sidebands are far detuned from both qubit types so that their effects can be neglected. Here we use focused $532\,$nm laser beams with a beam waist radius of about $3.5\,\mu$m, smaller than the ion spacing of about $9.5\,\mu$m. We use acousto-optic deflectors (AODs) to orient the laser beams to the two target ions, and we use symmetric optical paths for the two Raman beams such that the frequency shift due to the AODs are cancelled on each target ion \cite{patent2021,li2023low,hou2024individually}.

\begin{figure}[!tbp]
\includegraphics[width=\linewidth]{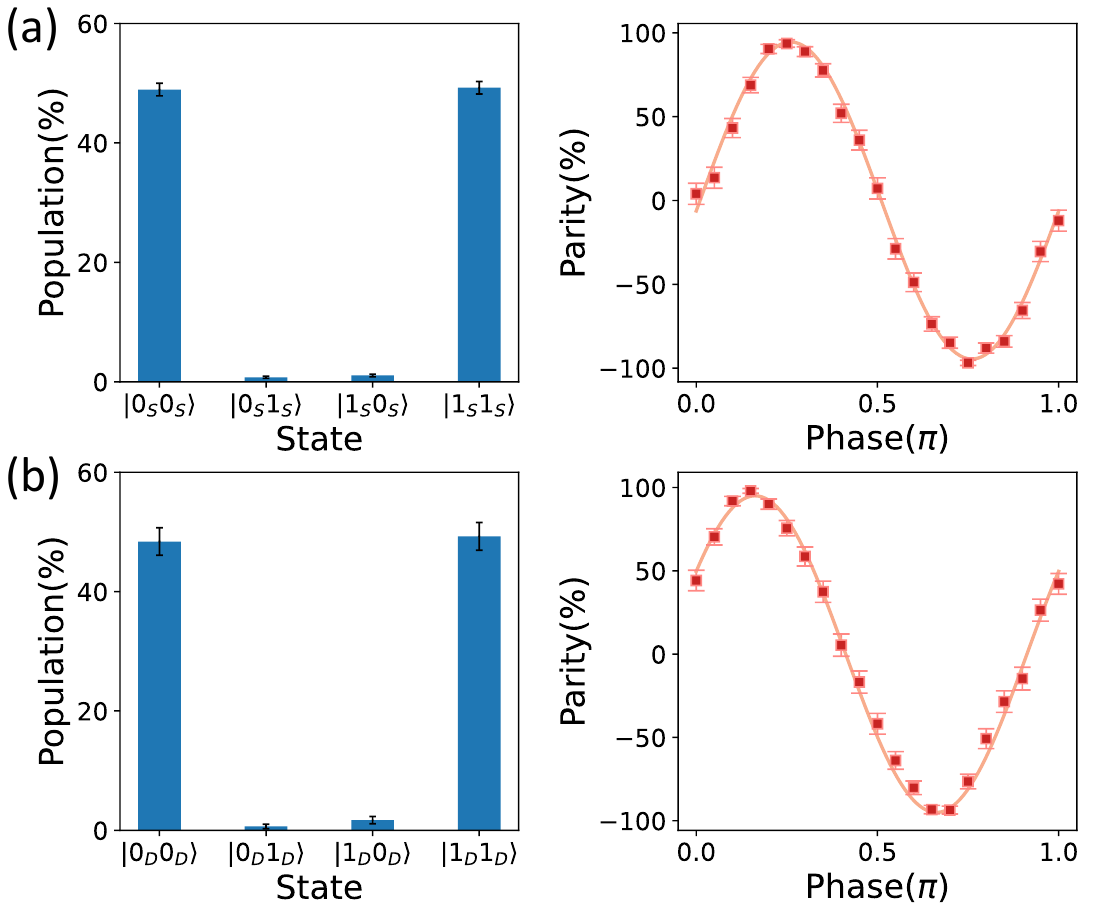}
\caption{Bell state fidelity for (a) $S$-type qubits and (b) $D$-type qubits after applying an entangling gate. The left panels represent the population of the prepared Bell state in the computational basis, while the right panels represent the parity oscillation. From these results, we obtain a fidelity of $96.4(4)\%$ for the $S$-type gate and $96.3(6)\%$ for the $D$-type gate. Each data point represents $500$ repetitions and the error bars correspond to one standard deviation.}
\label{fig:2}
\end{figure}

The $532\,$nm laser leads to differential AC Stark shift on both $S$-type and $D$-type qubits, which is difficult to suppress simultaneously by using a simple polarization of the laser. Therefore, here we choose to set the laser polarization to minimize the differential AC Stark shift for the $D$-type qubit as measured by the radio antenna. (A similar experiment has recently been demonstrated to minimize the AC Stark shift for the $S$-qubit of ${}^{133}\mathrm{Ba}^+$ \cite{PhysRevLett.132.263201}.) The remaining differential AC Stark shift on the $S$-qubit can be compensated during the entangling gate by fine-tuning the frequency components of the Raman laser beams via the EOM driving frequency. Next we calibrate the laser intensity for addressing the $S$-type and $D$-type qubits. Experimentally we measure the Raman transition Rabi rate of the $D$-qubit to be about $40\%$ of that of the $S$-qubit under the same laser intensity. In principle, for the dual-type entangling gate, it is not necessary to have the same Rabi rates for the two qubit types, as they only contribute to the overall two-qubit phase in the form of their product \cite{Lee_2005}. Nevertheless, in this experiment since we demonstrate the $S$-$S$, $D$-$D$ and $S$-$D$ gates using the same two-ion crystal, it is convenient to set equal Rabi rates for both qubit types via the EOM driving amplitude, so that all these gates can be achieved using the same gate parameters.

\begin{figure}[!tbp]
\includegraphics[width=\linewidth]{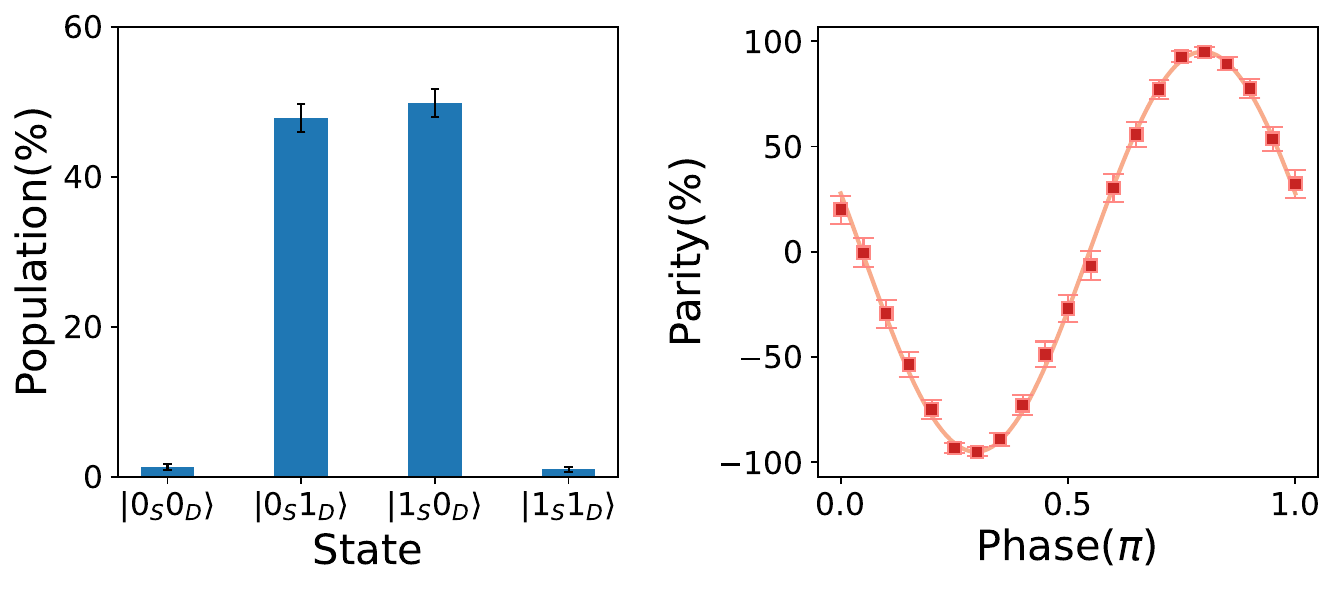}
\caption{Bell state fidelity for an $S$-$D$ qubit pair after a direct dual-type entangling gate.
A gate fidelity of $96.3(4)\%$ is obtained from the population of $97.7(5)\%$ (left panel) and the parity contrast of $95.0(3)\%$ (right panel). Each data point represents $500$ repetitions and the error bars correspond to one standard deviation.}
\label{fig:3}
\end{figure}

As shown in Fig.~\ref{fig:1}(a), in our setup, the Raman laser beams are at $45^\circ$ to both the $x$ and $y$ transverse modes, either of which can be used for the realization of the entangling gate. Due to the environmental noises, our $y$ modes have higher heating rates than the $x$ modes \cite{huang2024electromagnetically}, hence we set the laser detuning $\mu$ close to the $x$ phonon modes. Specifically, the center-of-mass mode and the rolling mode are at the frequencies of $\omega_c=2\pi\times 1.601\,$MHz and $\omega_r =2\pi\times 1.582\,$MHz, respectively. We set the laser detuning in the middle of these two modes as $\mu=(\omega_c+\omega_r)/2$, so that the two modes can be disentangled from the spin states simultaneously after one loop in the phase space with an evolution time of $T=4\pi/|\omega_c-\omega_r|=105.3\,\mu$s. By calibrating an appropriate overall laser intensity, a maximally entangled Molmer-Sorensen gate can thus be achieved \cite{Lee_2005}.

First we demonstrate the entangling gate for a same-type $S$-$S$ ion pair. We initialize both ions in $|0_S\rangle$ and apply an entangling gate. To evaluate the fidelity of the prepared Bell state, we measure the population of the final state in the computational basis and get a probability of $98.2(3)\%$ to be in the state $|0_S 0_S\rangle$ or $|1_S 1_S\rangle$ as shown in Fig.~\ref{fig:2}(a). Also we apply a global Raman $\pi/2$ rotation $R_\phi(\pi/2)$ to both qubits and measure the parity $p_{00}+p_{11}-p_{01}-p_{10}$ versus the phase $\phi$ to fit a parity contrast of $94.5(5)\%$. Combining these results, we obtain a Bell state fidelity of $96.4(4)\%$ for the $S$-$S$ pair \cite{debnath2016programmable}. As we shown in Supplemental Material, it is currently limited by the laser dephasing time of $2.6(2)\,$ms and the motional dephasing time of $4.1(5)\,$ms.

Next we consider a same-type $D$-$D$ ion pair. Here a subtle difference is the order in which we post-select the successful preparation of the initial $|0_D 0_D\rangle$ state. Previously we mention that we can verify the initial state to enhance the preparation fidelity. However, during the verification (state detection) stage, the ions are subjected to heating from the environment as well as the random scattering of the photons. On the other hand, when we have both ions in the $D$-type, after the verification step there will be no ancilla ions for sympathetic cooling, leaving the motional states at a higher temperature and thus reducing the gate fidelity. To avert this problem, we immediately apply the $D$-$D$ entangling gate after each attempt of preparing $|0_D 0_D\rangle$, and then verify that the ions stay in the $D_{5/2}$ subspace by the $493\,$nm and $650\,$nm lasers. For the post-selected dark events, we further measure a population of $97.6(7)\%$ and a parity contrast of $95.0(5)\%$ as shown in Fig.~\ref{fig:2}(b), hence a Bell state fidelity of $96.3(6)\%$. Note that in this way we exclude the leakage of the $D_{5/2}$ state from the gate infidelity, which is nevertheless expected to be small because the gate time of about $100\,\mu$s is much shorter than the lifetime of the $D_{5/2}$ levels of about $26\,$s \cite{PhysRevA.97.032508}. Also note that this post-selection will not be a limitation for the large-scale applications in the future because we can avoid this by performing sympathetic cooling with some ancilla ions in the $S$ state.

Finally we perform the dual-type $S$-$D$ entangling gate. We start from the two ions in $|0_S 0_S\rangle$. Then we use an individual $532\,$nm laser to create an AC Stark shift on one ion and apply a global $1762\,$nm laser to transfer the other ion to $|1_D\rangle$. After verifying the prepared $D$-qubit, we further perform sympathetic EIT cooling on the remaining $S$-qubit to cool down the whole system, and then reset this $S$-qubit into $|0_S\rangle$. Different from the previous cases, here we choose the initial state $|0_S 1_D\rangle$ rather than $|0_S 0_D\rangle$ for the convenience of measuring the parity oscillation of the generated Bell state. Due to our Raman laser configuration in Fig.~\ref{fig:1}(c), when scanning the relative phase between the counter-propagating Raman laser beams, the directions of the $R_\phi(\pi/2)$ rotations change oppositely for the $S$-type and $D$-type qubits. Therefore by flipping one qubit, we effectively reverse its phase scan into the same direction as the other qubit. As shown in Fig.~\ref{fig:3}, we obtain a population of $97.7(5)\%$ and parity contrast of $95.0(3)\%$, thus a Bell state fidelity of $96.3(4)\%$.

To sum up, we demonstrate the direct entanglement between dual-type qubits encoded on $^{137}\mathrm{Ba}^{+}$ ions. We achieve a fidelity of $96.3\%$ for the dual-type entangling gate, identical to the same-type entangling gates within the error bars. Therefore it will be advantageous to use the direct dual-type entangling gate and reduce the overhead for qubit type conversion. The current gate errors are dominated by the laser and motional dephasing times of a few milliseconds and can be improved in future upgrades with better stabilization of the optical paths and the locking of the trap frequency. The demonstrated direct dual-type entangling gate can find wide applications in quantum information sciences, reducing the circuit depth for quantum error correction and trapped-ion-based quantum network.

\begin{acknowledgments}
This work was supported by Innovation Program for Quantum Science and Technology (2021ZD0301601), Tsinghua University Initiative Scientific Research Program, and the Ministry of Education of China. L.M.D. acknowledges in addition support from the New Cornerstone Science Foundation through the New Cornerstone Investigator Program. Y.K.W. acknowledges in addition support from Tsinghua University Dushi program.
\end{acknowledgments}

\bibliography{reference}

\end{document}


\title{Supplemental Material for ``Experimental realization of direct entangling gates between dual-type qubits''}

\date{\today}

\maketitle

\section{Choice of the working magnetic field}
\begin{figure}[htbp]
\includegraphics[width=0.5\textwidth]{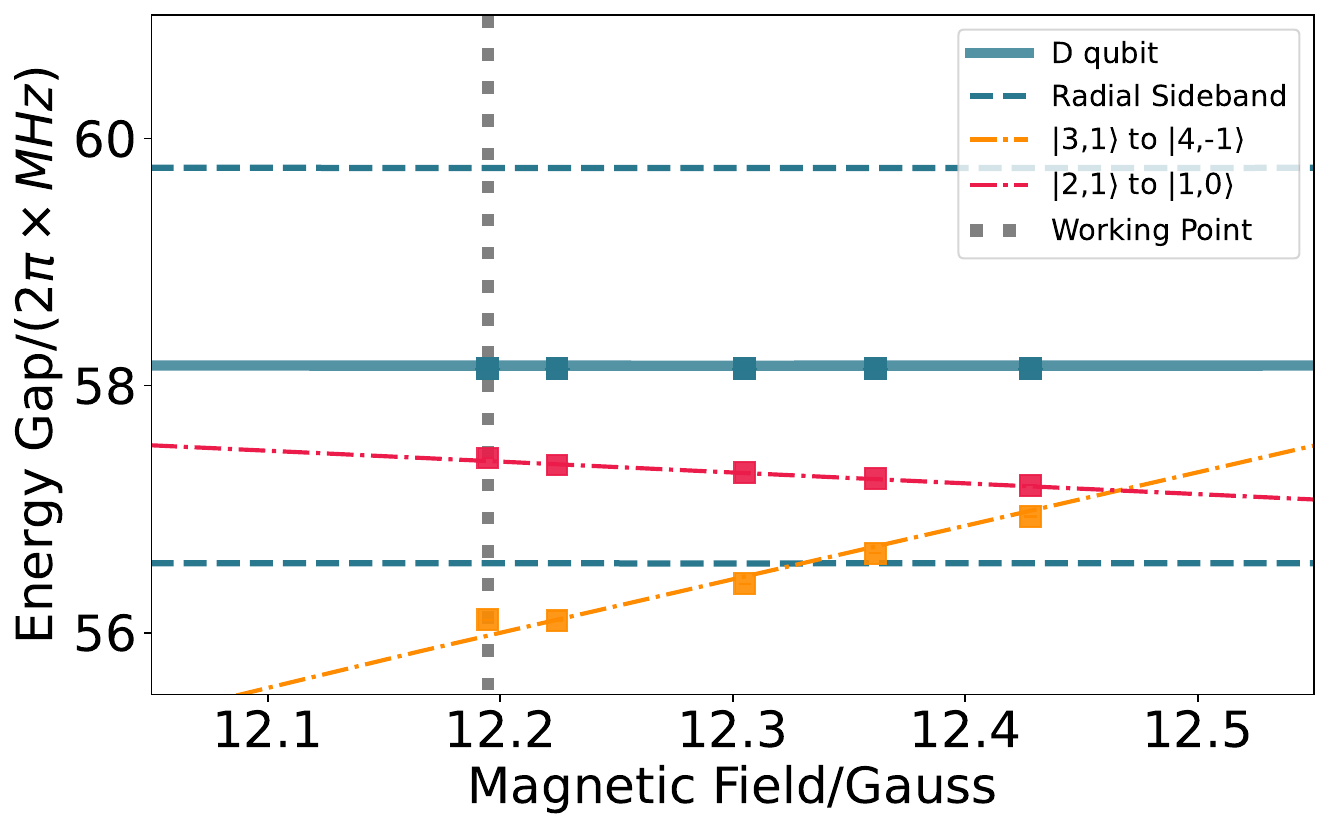}
\caption{Transition frequencies to the nearby $D_{5/2}$ manifold levels and the motional sidebands around the $D$-qubit frequency. To avoid off-resonant excitations, we choose a working magnetic field of $12.2\,$G.}
\label{fig:detuning}
\end{figure}

As described in the main text, we encode the dual-type qubits in $|0_{S}\rangle\equiv|S_{1/2},F=1,m_{F}=0\rangle$, $|1_{S}\rangle\equiv|S_{1/2},F=2,m_{F}=0\rangle$, $|0_{D}\rangle\equiv|D_{5/2},F=2,m_{F}=1\rangle$ and $|1_{D}\rangle\equiv|D_{5/2},F=3,m_{F}=1\rangle$, whose transition frequencies are insensitive to the small perturbation in the magnetic field at the working point of $B=12.2\,$G. Apart from the sensitivity to noise, a major consideration in choosing the working magnetic field is to avoid the off-resonant excitation to the nearby levels of the $D_{5/2}$ manifold or to their motional sidebands. In Fig.~\ref{fig:detuning} we plot the transition frequencies to the relevant levels, including the $D$-qubit frequency (blue solid), $|D_{5/2},F=2,m_{F}=1\rangle \to |D_{5/2},F=1,m_{F}=0\rangle$ (red dash-dotted), $|D_{5/2},F=3,m_{F}=1\rangle \to |D_{5/2},F=4,m_{F}=-1\rangle$ (yellow dash-dotted) and the motional sidebands of the radial phonon modes around $1.6\,$MHz (blue dashed). To suppress the off-resonant excitations, we want to avoid the crossing between the red and the yellow curves to the blue ones (including the solid curve for the carrier transition and the dashed ones for the motional sidebands). Therefore we choose a working magnetic field of $12.2\,$G where the detunings to the other transitions are above $0.5\,$MHz and the total off-resonant excitation is estimated to be about $0.1\%$.

\section{Measurement of laser and motional coherence time}
\begin{figure}[htbp]
\includegraphics[width=\textwidth]{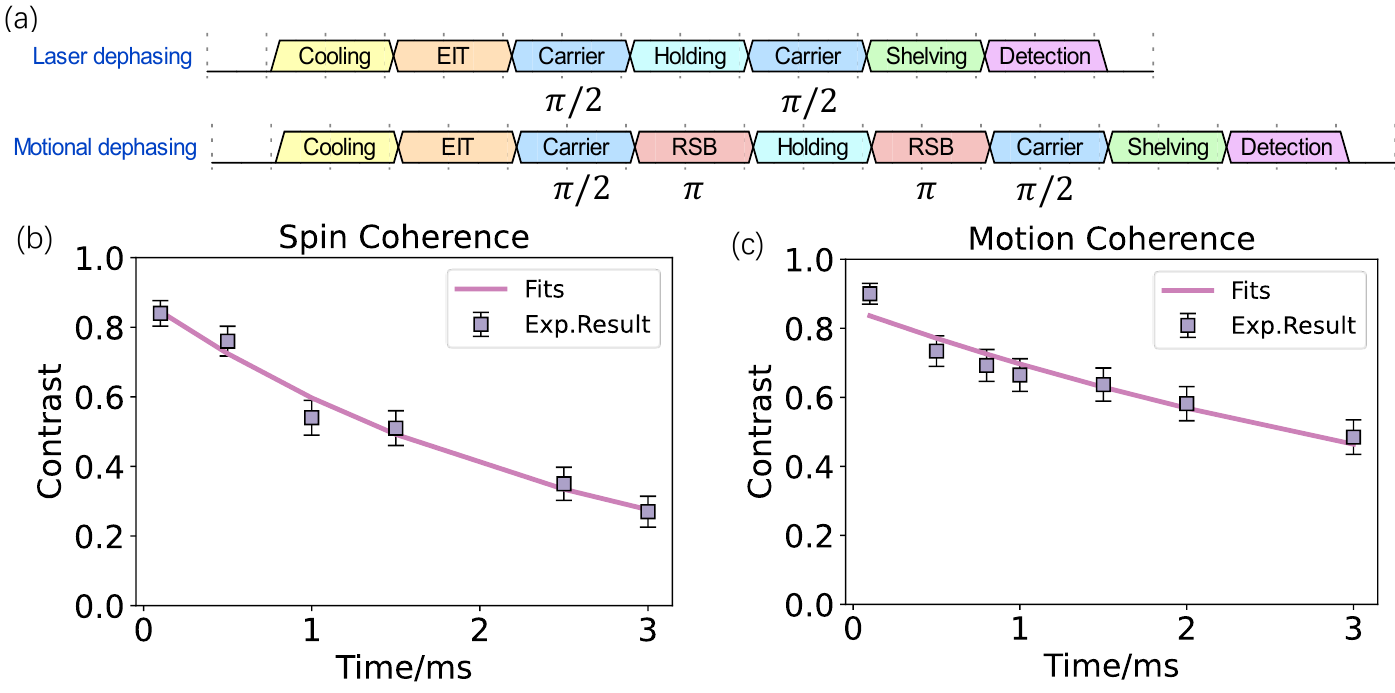}
\caption{(a). The sequence to measure the spin and motion phase coherence time. (b). Spin phase coherence is fitted as 2.59(0.22)ms. (c). Motion phase coherence is fitted as 4.92(0.69)ms.}
\label{fig:coherence}
\end{figure}

The experimental sequences to measure the laser and the motional dephasing time are plotted in Fig.~\ref{fig:coherence}(a). We measure the Ramsey fringes under the counter-propagating Raman laser beams to obtain the spin dephasing time which is dominated by the fluctuation in the optical path of the laser. We further combine a carrier $\pi/2$ pulse and a red sideband $\pi$ pulse to cancel the optical phase and to measure the motional dephasing time \cite{hou2024individually}. By fitting the exponential decay as shown in Figs.~\ref{fig:coherence}(b) and (c), we get a laser dephasing time $\tau_s=2.6(2)\,$ms and a motional dephasing time $\tau_m=4.1(5)\,$ms, respectively.

\section{Experimental pulse sequence}
\begin{figure}[htbp]
\includegraphics[width=\textwidth]{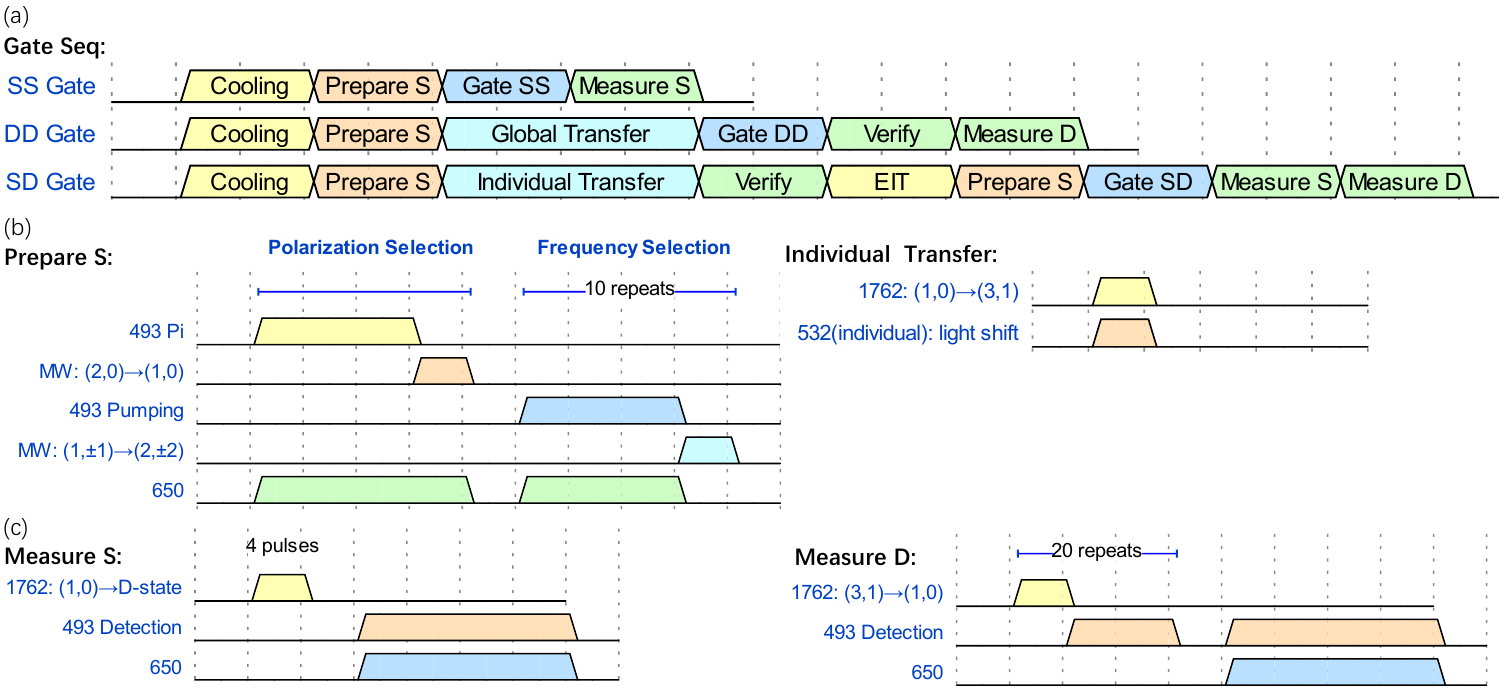}
\caption{The experimental sequences for the entangling gates between the same-type and dual-type qubits (a), and the corresponding state preparation (b) and measurement (c) processes.}
\label{fig:sequence}
\end{figure}

The experimental sequences for the state initialization, readout and the entangling gates are summarized in Fig.~\ref{fig:sequence}. The detailed explanations can be found in the main text. We use $|D_{5/2},F=2,m_{F}=0\rangle$, $|D_{5/2},F=1,m_{F}=0\rangle$, $|D_{5/2},F=1,m_{F}=1\rangle$ and $|D_{5/2},F=4,m_{F}=1\rangle$ for the electron shelving of the $S$-qubit.

\section{Error Budget}
The error budget for the $S$-$D$ dual-type entangling gate is shown in Table~\ref{tab:budget}. The laser dephasing, motional dephasing and motional heating errors are estimated by a Lindblad master equation using the measured dephasing times and the heating rate. The off-resonant excitation error is estimated from the calculated detuning and the measured Rabi rates. The SPAM errors are estimated individually from the daily operations of the trap.
\begin{table}[htbp]
    \centering
    \caption{Error budget for the dual-type entangling gate.}\label{tab:budget}
    \begin{tabular}{|c|c|}
        \hline
        \textbf{Error Source} & \textbf{Gate Error(\%)}  \\ \hline
        Laser dephasing & $1.8(1)$  \\
        Motional dephasing & $1.1(2)$  \\
        Motional heating  & $0.4(1)$  \\
        Off-resonant excitation & $0.1$  \\
        SPAM error & $<0.3$ \\
        \hline
    \end{tabular}
\end{table}

\bibliography{reference}